\documentclass[aps,prf,preprint,amsmath,amssymb,superscriptaddress]{revtex4-2}

\usepackage{graphicx}
\usepackage{dcolumn}
\usepackage{bm}
\usepackage[colorlinks,citecolor=blue]{hyperref}
\usepackage{mathrsfs}
\usepackage{subfig}
\usepackage{graphicx}
\usepackage{epstopdf}
\usepackage{natbib}
\usepackage{siunitx}
\usepackage{booktabs}
\usepackage{xcolor}
\usepackage{cleveref}
\begin{document}


\title{Mixed mode transition in boundary layers: Helical  instability}



\author{Rikhi Bose}
\email{rikhi.bose@gmail.com}
\affiliation{Max Planck Institute for Solar System Research, G\"{o}ttingen, 37077, Germany}
\affiliation{Engineering Laboratory, National Institute of Standards \& Technology, Gaithersburg, MD 20899, USA}
\author{Paul A. Durbin}
\affiliation{Aerospace Engineering, Iowa State University, Ames, Iowa 50011, USA}


\date{\today}

\begin{abstract}
Recent  (Bose \& Durbin, \textit{Phys. Rev. Fluids}, 1, 073602, 2016) direct numerical simulations (DNS) of adverse- and zero-pressure-gradient boundary layers beneath moderate levels of free stream turbulence ($Tu$ $\le 2\%$) revealed a \textit{mixed mode} transition regime, intermediate between orderly and bypass routes. 
In this regime, the amplitudes of the Klebanoff streaks and instability waves are similar, and potentially can interact. 
Three-dimensional visualizations of transitional eddies revealed a helical pattern, quite distinct from the sinuous and varicose forms seen in pure bypass transition. 
This raises the fundamental question of whether the helical pattern could be attributed to a previously unknown instability mode.

Two-dimensional stability analyses are performed herein for base flows extracted from DNS flow fields. 
The three-dimensional structure of the eigenfunction of the most unstable mode indeed reveals a helical pattern.
These instability modes are obtained significantly earlier than 
helical structures are seen in flow visualizations, even for innocuous base flow structures.  It is inferred that the instability mode is the underlying cause of helical
patterns that emerge in the transition region.
The streak configuration leading to the formation of the helical mode instability is different from 
those leading to  sinuous and varicose modes in previous studies of pure bypass transition. 
Thus, the mixed mode precursor is the distinctive cause. 
\end{abstract}


\maketitle

\section{Introduction}
\label{intro}

Direct numerical simulations (DNS) have provided a detailed description of the laminar to turbulent transition process in attached subsonic boundary layers. 
Both orderly  and bypass transition have been investigated in detail. 
The theory of orderly transition is well established.  It starts with exponential growth of instability waves, followed by a much faster secondary instability process forming $\Lambda$-vortices. 
Finally, lifted $\Lambda$-vortices  break 
down into patches of turbulence --- that final stage is impervious to theory. 
When transition does not emerge from growing instability waves it is classified as bypass transition. 
In most of the literature,  bypass transition is synonymous with transition due to incident free-stream turbulence (FST). 
Over time various theoretical concepts have emerged  \citep{durbin2017perspectives}.  They explain the occurence of
streamwise elongated perturbations, called Klebanoff streaks, as precursors to transition.  These disturbances
amplify via the lift up mechanism, or algebraic growth,  
until their amplitudes saturate.  Klebanoff streaks undergo secondary instability.  The nature of that secondary instability is less
clear cut than the $\Lambda$-vortices of orderly transition.  Analysis have characterized it in terms of  sinuous and varicose modes,
and outer and inner instabilities \citep{Andersson2001, hack_zaki}. 

Experimental evidence suggests the existence of flow regimes 
in which both instability waves and streaks are involved in the transition process \citep{walker_gostelow, Westin}. 
Especially, this can occur in adverse pressure gradient (APG) under low to moderate levels of free-stream turbulence. 

A recent DNS study \citep{bose2016helical, bose2016transition,bose2018instability}  revealed an intriguing transition process: 
beneath moderate levels of FST (intensity $Tu \le 2\%$), and in the 
presence of instability waves at super-critical Reynolds numbers, a truly \textit{mixed mode} transition 
process was reported. 
Three-dimensional visualizations of the perturbation fields revealed a features quite distinct from the 
sinuous and varicose modes of pure bypass transition. 
The flow structures resembled a helical pattern. 
A deep lying critical layer was reported for these helical eddies, and therefore, these were classified as \textit{inner} 
instabilities---in the
terminology of \citet{vaughan_zaki}---but, again, distinct from what had been seen previously. 

Those empirical observations pose the fundamental questions, {\it does a helical, secondary streak instability exist?} 
and,  {\it is it the cause of the patterns seen in flow visualization?} Those questions prompted the present study.  We analyze the  
secondary stability of boundary layers in the mixed mode transition regime. 
Both zero and APG boundary layers are considered.   The approach parallels that of \citet{hack_zaki}: 
observed helical patterns are tracked back in time to  identify their precursors.
These provide base flow fields to which stability analysis is applied. 
An iterative technique is used to extract those eigenmodes having the highest growth rates. 

As will be shown, the eigenfunction solution that is obtained is indeed of helical form.  These eigenmodes are obtained even for apparently innocuous base streaks, significantly earlier than the helical instability is seen in DNS.  This substantiates
the hypothesis that the helices that were seen in computer simulations can be attributed to a new type of 
instability.

The streak configuration leading to the inner helical mode is distinct from the precursors to previous solutions of
secondary instabilities. 
Outer sinuous and inner varicose streak instabilities have been attributed to the inflection points in the spanwise and
 wall-normal profiles of the streamwise velocity, respectively  \citep{Andersson2001, hack_zaki}.
The inner helical mode appears at the intersection of low- and high-speed streaks, where spanwise and wall-normal shear 
are both important. 
Such unique orientation of the streaks is due to the high growth rate of the mixed mode precursors, formed as a consequence of the instability waves interacting with the Klebanoff streaks.  As is inevitable, the new instability, discovered herein, is a 
consequence of the precursor differing from previous studies.

\section{Numerical techniques}\label{methods}

\subsection{Direct Numerical Simulations}\label{dns}

The direct simulations were performed in the mixed mode transition regime for both ZPG and APG boundary layers in order
to create the precursor fields.
A detailed description of the numerical methods, boundary conditions, and the method for the generation of inlet turbulence can be found in \citep{bose2016mixed, bose2018instability} and are not repeated here.
For all simulations,  inlet  FST with a prescribed intensity ($Tu \equiv \sqrt{\frac{1}{3}\overline{u'_i u'_i}}/U_\infty  = 1\%$ and $2\%$) 
is superimposed on a Blasius boundary layer
\citep{Jacobs_Durbin, BrandtDNS} at a super-critical Reynolds number, $Re_b = \sqrt{U_\infty x_0/ \nu} \approx 398$. 
In ZPG simulations, the inlet additionally includes a TS-wave with amplitude $A=0.5\% U_\infty$. 
The domain is rectangular for the ZPG simulations with size $320 \delta_0 \times 24 \delta_0 \times 24 \delta_0$ ($\delta_0$is the boundary-layer thickness at inlet) in the streamwise, wallnormal, and spanwise directions.
The grid has $1,537\times 161 \times 241$ points. 
The computational domain and grid specifications for the APG simulations are the same as the ZPG simulations except that the APG is induced by a curved upper boundary characterized by the Hartree parameter, $\beta_H=-0.14$ \citep{bose2018instability}. 
The largest grid spacings in viscous units are reported in table \ref{t1}. 
Despite the use of different grids, the mean skin-friction curves from the present simulations are in good agreement with those from previous studies \citep{bose2016transition, bose2016mixed, bose2018instability}. 
Once a statistically stationary state was reached in the simulations, instantaneous snapshots were stored
 at constant time intervals of $1.8\delta_0/U_\infty$, over a period spanning at least three flow-through times,  for stability analysis.

\begin{table}
	\centering
	\begin{tabular*}{0.75\linewidth}{@{\extracolsep{\fill}}c c c c c}
        \hline
        PG  &  $Tu$           &  $\Delta x^+_{max}$ &  $\Delta y^+_{1, max}$ &  $\Delta z^+_{max}$ \\
        \hline		
        APG &  $1\%U_\infty$  &          22.58      &     1.198              &     10.83           \\
        APG &  $2\%U_\infty$  &          22.35      &     1.172              &     10.73           \\      
        ZPG &  $1\%U_\infty$  &          20.71      &     0.97               &      9.94           \\      
        ZPG &  $2\%U_\infty$  &          20.4       &     0.95               &      9.79           \\   
	\hline
	\end{tabular*}
	\caption{DNS: Largest grid spacings in viscous units}
	\label{t1}
\end{table}

\subsection{Stability analysis}\label{methods:stability}

Base flow profiles are extracted in  planes normal to the streamwise direction, from instantaneous DNS fields.  Only the streamwise component of the velocity field is included,
 $[\boldsymbol{U}_b, p_b]=[\{U_b(y, z; x, t), 0, 0\}, 0]$.  The $x,t$ dependence is treated as parametric for the stability analysis. 
Perturbations are periodic in the streamwise direction,  of the form
\[
[\boldsymbol{u}_2', p_2'](x, y, z, t)=[\{\hat{u}_2, \hat{v}_2, \hat{w}_2\}, \hat{p}_2](y, z, t)e^{i k_x x},
\]
 where, $k_x$ is the streamwise wavenumber, also extracted from the DNS. 
These idealizations can be formally justified, as the streaky base flow is predominantly in the streamwise direction, and changes 
far more slowly than the growth rate of the perturbations in  $x$~and~$t$ \citep{Andersson2001, hack_zaki}. 
Consequently, the following set of coupled linear equations is obtained for the perturbations, 
\begin{eqnarray}
\label{mod:cont}
 &(i k_x) \hat{u}_2  + \frac{\partial \hat{v}_2}{\partial y} + \frac{\partial \hat{w}_2}{\partial z} = 0  \\
\nonumber
 \frac{\partial \hat{u}_2}{\partial t} &= \big[-ik_x U_b + \Delta \big]\hat{u}_2 + \big[ -\frac{\partial U_b}{\partial y}\big]\hat{v}_2 + \big[-\frac{\partial U_b}{\partial z}\big]\hat{w}_2 + \big[ -ik \big]\hat{p}_2	\\
\label{mod:w}
 \frac{\partial \hat{v}_2}{\partial t} &= \big[ 0 \big]\hat{u}_2 + \big[ -ik_x U_b + \Delta \big]\hat{v}_2 + \big[0 \big]\hat{w}_2 + \big[ -\frac{\partial}{\partial y} \big]\hat{p}_2 \hskip4em\\
\nonumber
 \frac{\partial \hat{w}_2}{\partial t} &= \big[ 0 \big]\hat{u}_2 + \big[ 0 \big]\hat{v}_2 + \big[-ik_xU_b + \Delta \big]\hat{w}_2 + \big[ -\frac{\partial}{\partial z} \big]\hat{p}_2\hskip4em	
\end{eqnarray}
The Laplacian is $\Delta = \frac{1}{Re}\big( -k_x^2 + \frac{\partial^2}{\partial y^2} + \frac{\partial^2}{\partial z^2} \big)$.  $\big[ 0 \big]$ indicates a zero in a matrix element of
the discretized equations.

The perturbations satisfy the following boundary conditions,
\begin{eqnarray}
\label{bc1}
& \hat{u}_2, \hat{v}_2, \hat{w}_2 = 0 \textrm{ at  } y = 0 \\
\label{bc2} 
& \frac{\partial \hat{u}_2}{\partial y}, \hat{v}_2, \frac{\partial \hat{w}_2}{\partial y} = 0 \textrm{ as  } y \to \infty	
\end{eqnarray}
As in the DNS, a periodic boundary condition is applied in the spanwise direction. 

Following \citet{barkley1996three}, the set of autonomous equations may be written as 
\begin{equation}
\label{bh}
\frac{\partial \hat{\boldsymbol{u}}_2}{\partial t} = \mathbf{C} \hat{\boldsymbol{u}}_2 
\end{equation}

Denoting $e^{\mathbf{C} t} = \mathbf{A}(t)$,
formal integration of (\ref{bh}) yields 
\begin{equation}
\label{hz2}
\hat{\boldsymbol{u}}_2(t)= \mathbf{A}(t)\hat{\boldsymbol{u}}_2(0) \equiv e^{-i \omega t} \hat{\boldsymbol{u}}_2(0)
\end{equation}
 The eigenspectrum of the matrix exponential, $\mathbf{A}(t)$, dictates the time evolution of the eigenfunctions of $\mathbf{A}(t)$. 
The real and imaginary parts of $\omega$ provide the frequency and growth rate of disturbances, respectively. 
Instead of an explicit computation of the eigenvalues of the large, two-dimensional stability problem, an iterative method is used following \citet{sorensen2002numerical}. 
In this method, eigenvalues with the highest growth rates, and also those that yield phase speeds similar to that extracted from DNS,
are extracted with the implicitly restarted Arnoldi method (IRAM).  The Francis QR factorization technique 
is used to extract eigenvalues.
A fuller description of the iterative scheme is provided in \citet{hack_zaki}, and is not repeated here. 

Formally, the iterative eigenvalue solver applies the linear vector differential operator $\mathbf{A}(t)$ repeatedly to an initial disturbance field as in (\ref{hz2}). 
In practice, this is equivalent to
repeatedly integrating (\ref{mod:cont}) and (\ref{mod:w}) in time, with the boundary conditions (\ref{bc1}) and (\ref{bc2}). 
Time integration is with a fractional  step method, similar to the solution scheme used for the DNS:
advection uses the second order Adams Bashforth scheme, and diffusion uses the implicit Crank Nicolson scheme \citep{bose2016transition}. 
The spatial discretizations in both wall-normal and spanwise directions is by second-order central differencing. 
For both the  momentum equation and the pressure correction steps, periodicity in the spanwise direction is exploited by solving for coefficients of  
spanwise Fourier modes, followed by an inverse transform. 
161 and 482 grid points are used to discretize the equations in the wall-normal and spanwise directions, respectively.

An important consideration is the calculation of the frequency of the instability mode (equation \ref{hz2})
\begin{equation}
\label{omg}
\omega_r \Delta t = -\theta 
\end{equation}
\noindent where, $\theta$ is the phase of a complex eigenvalue of $\mathbf{A}(t)$, and $\Delta t$ is the interval of time integration. 
The iterative scheme is based on most unstable mode dominating at large times. 
Care is needed in choosing $\Delta t$, as an inverse trigonometric function is required to compute the modal frequency. 
$\Delta t$ was chosen in conjunction with applying $\mathbf{A}(t)$ multiple times, to ensure $|\omega_r \Delta t| \sim \pi / 4$ which is within the  range of the 
inverse trigonometric function used. 

The instantaneous base flow, from the DNS, incorporates the full non-linear evolution of the boundary layer populated by instability waves and Klebanoff streaks. 
A series of base states is extracted from DNS by streamwise displacement of the sampling plane at the phase speed of the disturbance observed in DNS. 
In this way, evolution of a helical disturbance from quasi-parallel stability analysis can be compared, in an approximate sense, to DNS. 

\section{Results \& Discussion}\label{results}

\begin{figure}
  \centerline{\includegraphics[trim=0 1cm 0 0,clip,width=0.8\textwidth]{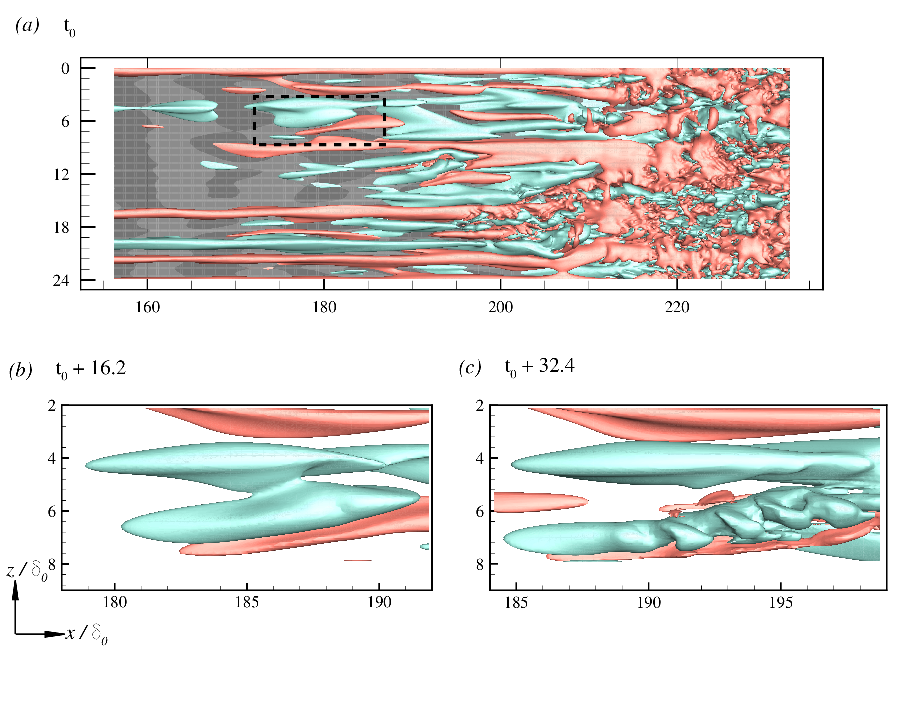}}
    \caption{Top view of the surfaces of iso-$u'$ (red and green indicate high- and low-speed regions, respectively): ($a$) $u' = \pm 0.06$; Contours of $-0.2\le v' \le 0.2$ (darker is negative) at a wall-parallel plane at $y = 0.5\delta_0$. ($b$) and ($c$) $u' = \pm0.1$.}
\label{u-iso}
\end{figure}

In the mixed mode transition regime, the primary flow field is a juxtaposition of Kelbanoff streaks and instability waves on the boundary
layers. 
In the APG simulations, the instability waves arise spontaneously \citep{bose2018instability}, while in ZPG, an unstable TS wave is injected at the inlet of the simulation domain. 

Figure~\ref{u-iso}($a$) depicts the primary disturbance field at three time instants, from the APG simulation with $Tu=1\%$. 
Instantaneous iso-surfaces of streamwise perturbation velocity, $u'$ are shown along with the contours of wall-normal perturbation velocity, $v'$, at a wall-parallel plane $y/\delta_0 = 0.5$. 
While the streaks are prominent in the iso-surfaces of $u$, the instability waves are identifiable in $v$. 

The streaks are locally distorted by the instability waves and form local patches of mixed mode precursors. 
The black dashed box encloses one such precursor. 
Once this precursor forms, it amplifies much faster than both the streaks and instability waves, 
spawning the helical instability, which breaks down quickly to form a turbulent spot. 
The development of the precursor, formation of the helical instability and breakdown take place over a short streamwise distance of
 $10-20\delta_0$ as shown in figure \ref{u-iso}($b$) and \ref{u-iso}($c$). 
The streamwise size of the precursor is $\sim 10\delta_0$ (similar to the wavelength of the instability wave), while the wavelength of the helical instability is $\sim 1.3$--$2\delta_0$. 
How the mixed mode precursor originates is beyond the scope of the current paper and will be the subject of a future article. 
 
\begin{figure}
  \centerline{\includegraphics[trim=0 0 0 0,clip,width=.9\textwidth]{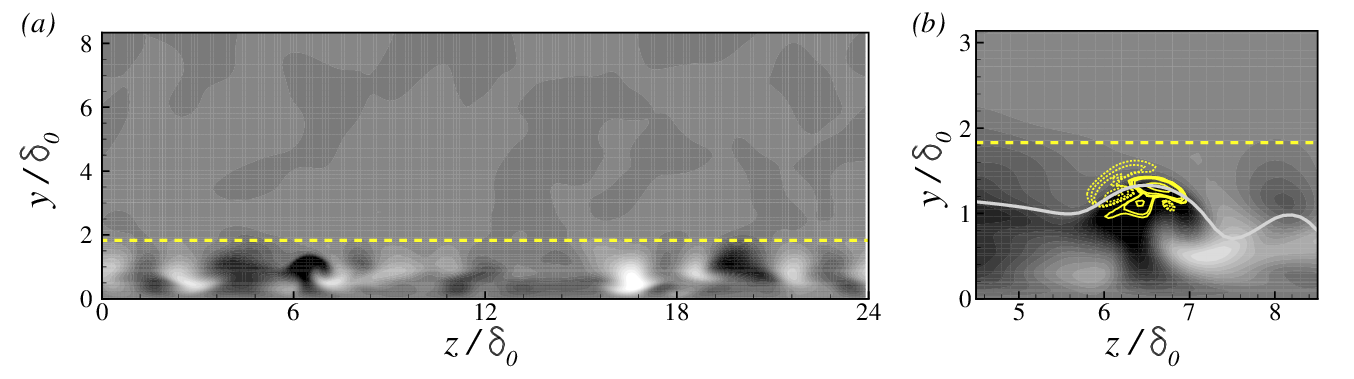}}
    \caption{Contours of $-0.2 \le u' \le 0.2$ (dark for negative values) at time $t_0 + 16.2$ as in figure \ref{u-iso}($b$) at the streamwise location $x/\delta_0 \sim 185.8$: ($a$) The full spanwise domain is shown; ($b$) Zoomed-in view of the helical mode precursor. The yellow contours show the real part of the streamwise perturbation eigenfunction (dashed lines indicate negative values). The
    light grey solid line and the horizontal yellow dashed line mark the local critical layer and $\delta_{99}$, respectively.}
\label{yz-plane}
\end{figure}

A $y, z$--plane extracted at $x/\delta_0\sim 185.8$ at the
same time instant, $t_0+16.2$, as figure \ref{u-iso}($b$) is shown in figure \ref{yz-plane}. 
Figure \ref{yz-plane}($a$) shows the full spanwise extent of the domain; a zoomed-in view of the
 precursor on which the helical instability develops is presented in figure \ref{yz-plane}($b$). 
Figure \ref{yz-plane}($b$) includes contours of the streamwise perturbation eigenfunction corresponding to the eigenmode with the highest growth rate computed via (\ref{mod:w}) for this base flow configuration. 

$k_x$ is an input parameter of the stability calculation.  It is 
estimated from the helical instability in the snapshot shown in figure \ref{u-iso}($c$). 
It is evident that the instability develops on a high-amplitude, mushroom-like, low-speed streaky structure. 
Unlike the inner varicose instability described by \citet{hack_zaki}, which forms when a high-speed streak climbs up a low-speed streak from behind, the inner helical instability forms when a low-speed  structure overlaps a high-speed structure from the side (also see figure 2 in \cite{bose2016helical}). 
The low phase speed, $c\approx0.546$ of this instability mode confirms that the helical mode is an inner instability. 
The critical layer for the instability is shown in figure \ref{yz-plane}($b$). 

$x, y$--planes through a helical mode \citep{bose2016helical, bose2016transition} reveal a Kelvin-Helmholtz type 
instability which is similar in description to the inner varicose mode. 
However, unlike the inner varicose modes \citep{hack_zaki},  figure \ref{yz-plane}(b) shows that the low-speed structure 
positions on top of the high-speed structure for a helical mode. 

\begin{figure}
  \centerline{\includegraphics[trim=0 0 0 0,clip,width=1\textwidth]{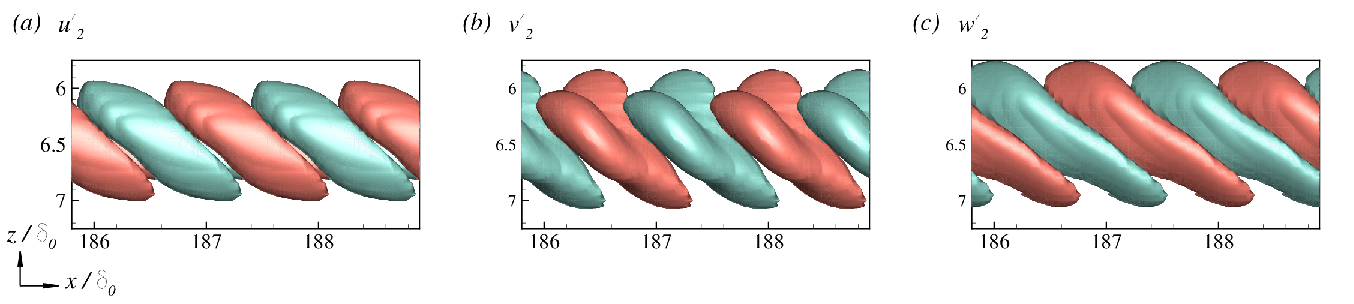}}
    \caption{Top view of the iso-surfaces of perturbation eigenfunctions (red and green represent positive and negative perturbation, respectively) for the most unstable mode from stability analysis plotted over two wavelengths: ($a$) $u_2' = \pm 0.02$; ($b$) $v_2' = \pm 0.015$ ($c$) $w_2' = \pm 0.02$.}
\label{eigfunc}
\end{figure}

Figure~\ref{eigfunc} is a three-dimensional view of  velocity components of the most unstable eigenmode, over two 
streamwise wavelengths.  This clearly reveals a helical pattern.
It  is quite distinct from the previously reported sinuous and varicose type instabilities: contrast to figure 20 of \citet{hack_zaki}.
 

The disturbance field from the stability calculations is in qualitative agreement with the helical mode observed in the DNS of figure \ref{u-iso}($c$). 
The chirality of the helix depends on the direction of the side-wise overlap of the low- and the high-speed streaks. 
For this particular instance of the helical mode, the overlap is from the right of the low-speed streak, and therefore, looking downstream, the helix wraps clockwise around the low-speed streak. 
In cases where it is on the left, the helical instability rotates counter-clockwise.

\begin{figure}
  \centerline{\includegraphics[trim=0 0 0 0,clip,width=.9\textwidth]{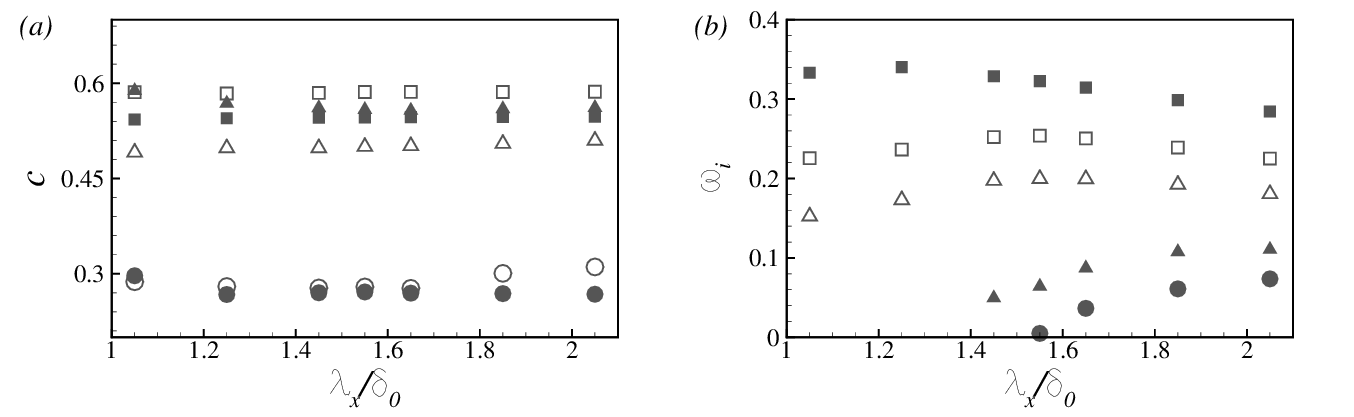}}
  \caption{($a$) Phase speed, $c$, and ($b$) growth rate, $\omega_i$, plotted against streamwise wavelength, $\lambda_x/\delta_0$, for the three most unstable modes for base states extracted from DNS at $t_0+16.2$ and $x/\delta_0\sim185.8$ (filled symbols), and at $x/\delta_0\sim190.2$ and $t_0+27$ (blank symbols): $\square$, most unstable, $\bigtriangleup$, second most unstable, and $\bigcirc$, third most unstable modes.} 
\label{modes}
\end{figure}

Although the most unstable helical eigenmode (hereafter called mode 1) is shown in figures \ref{yz-plane} and \ref{eigfunc}, the stability analysis predicts multiple unstable modes for base states like that shown in figure \ref{yz-plane}. 
The phase speed, $c$, and the temporal growth rate, $\omega_i$, for the three most unstable modes of
 the two base states extracted at
 time instants $t_0+16.2$ (figure \ref{yz-plane}) and $t_0+27$ (at this instant, the helical instability is about to become prominent in DNS) are plotted against the streamwise perturbation wavelength, $\lambda_x/\delta_0$, in figure \ref{modes}. 
The second most unstable mode (mode 2) has phase speed similar to mode 1. 
However, its growth rate is lower for all $\lambda_x/\delta_0$. 
For mode 1, highest value of $\omega_i$ is obtained for $\lambda_x/\delta_0 \sim 1.25$ at $t_0+16.2$, and for $\lambda_x/\delta_0 \sim 1.55$ at $t_0+27$. 
Therefore, the stability analysis predicts an increase in the streamwise wavelength of the helical instability with downstream distance.
This is qualitatively consistent with the DNS. 
The third most unstable mode (mode 3) has a significantly lower phase speed than modes 1 and 2, and is stable for all $\lambda_x/\delta_0$ at inception of the helical instability, $t_0+27$. 
At the earlier instant,  $t_0+16.2$, this mode is unstable for $\lambda_x/\delta_0 > 1.5$. 
The mode shapes for modes 2 and 3 are presented in appendix \ref{appA}. 

\begin{figure}
  \centerline{\includegraphics[trim=0 0 0 0,clip,width=.9\textwidth]{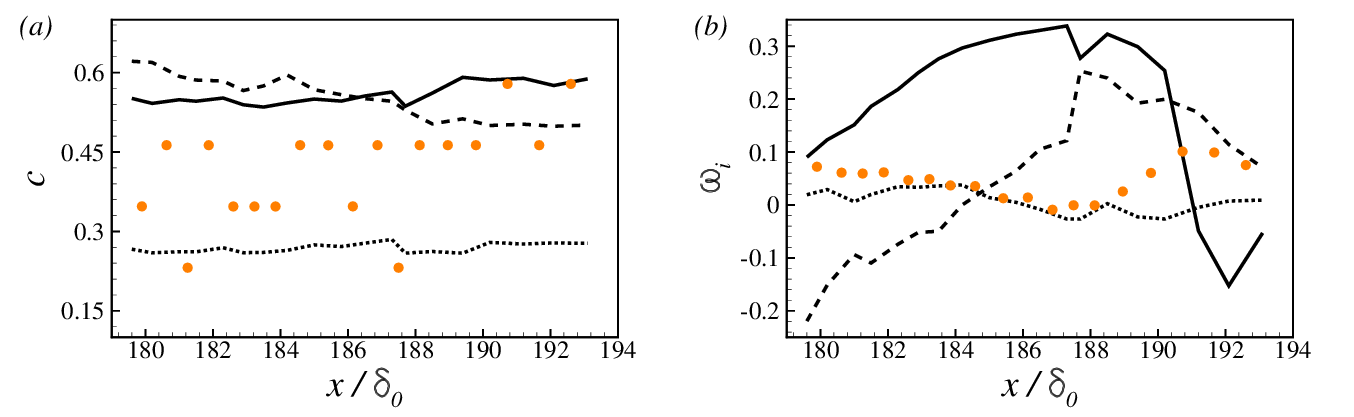}}
  \caption{Streamwise evolution of a helical instability from DNS (symbols) and stability analysis,
    ---$\!$---, mode 1; - - - - -, mode 2; $\cdots$\,$\cdots$, mode 3: ($a$) Phase speed; ($b$) Growth rate.}
\label{track}
\end{figure}

The stability analysis was performed on a series of base states extracted from DNS following the evolution of the helical mode. 
The streamwise wavelength for the stability analysis is $\lambda_x/\delta_0 = 1.55$, for which $\omega_i$ was highest at $x/\delta_0 \sim 185.8$ and $t = t_0+16.2$ (see figure \ref{modes}), and is approximately the streamwise average wavelength of the helical mode in DNS. 
The results (phase speed and growth rate) are compared with those extracted from DNS in figure \ref{track}. 
Phase speed is extracted from the DNS data by following the helical mode precursor backward in time and applying a 
finite difference formula. 
The spatio-temporal growth rate is extracted by following the peak of $v'$ corresponding to the instability in a wall-parallel plane 
located at $y/\delta_0 = 1$ (the plane is close to the critical layer obtained for the inner helical mode shown in figure \ref{yz-plane}) 
and using the formula
\[
\omega^i_{DNS}=\frac{1}{\Delta t} \ln\left|\frac{v'(t'+\Delta t)}{v'(t')}\right|.
\]

In figure \ref{track}($a$), phase speeds extracted from DNS show four discrete levels, indicating a superposition of instabilities propagating downstream. 
The phase speed of the unstable modes from stability analysis remains approximately constant with downstream distance. 

The superposition of the unstable modes obtained from stability analysis has some elements of consistency with the DNS. 
The growth rate, $\omega_i$ for the helical mode is significantly larger than those extracted from DNS. 
The temporal growth rate of mode 3 is in good agreement with the DNS; but, 
this mode has a smaller phase speed than those extracted from the DNS. 
Mode 2 is initially stable, and only becomes unstable at $x/ \delta_0 \sim 184.2$. 
Note that, as shown in figure \ref{yz-plane}, the helical mode is the most unstable well before it becomes
clearly seen, at $t_0+27$ and $x/\delta_0 \approx 192$. 
Its high growth rate means that the time interval between its inception and breakdown is very short. 
In DNS's, the breakdown of the helical mode after its genesis is also very short; it can only be clearly identified in iso-surface plots at $t_0+27$, and it breaks down at $\sim t_0+36$. 

In figure \ref{track}, the growth rate of the most unstable helical mode does not match with those extracted from DNS. 
There are two reasons: firstly, the stability analysis is linear,  disturbances are monochromatic, and the growth rate is only temporal. 
The DNS is non-linear and the growth rate is spatio-temporal.
Secondly, and perhaps more importantly, the length and time scales of the helical instability are an order of magnitude smaller than the base-state instability waves (the scales of the Klebanoff streaks are even larger). 
Therefore, at the beginning of the transition region, the scales corresponding to the helical instability have negligible energy. 
Once they become unstable, their energy grows at exponential rate. 
Hence, although the helical mode is the most unstable, at first, the primary instability has most of the disturbance energy
and dominates the growth rate observed in the DNS. 
As the disturbance energy at the helical-mode scales become of similar magnitude as the primary instability, they show up in the 
flow fields, but quickly break down to turbulence. 
This has been previously verified by plotting spectra,, e.g., figure 17 in \cite{bose2018instability}. 

\subsection{Helical mode in ZPG}

\begin{figure}
  \centerline{\includegraphics[trim=0 0 0 0,clip,width=.9\textwidth]{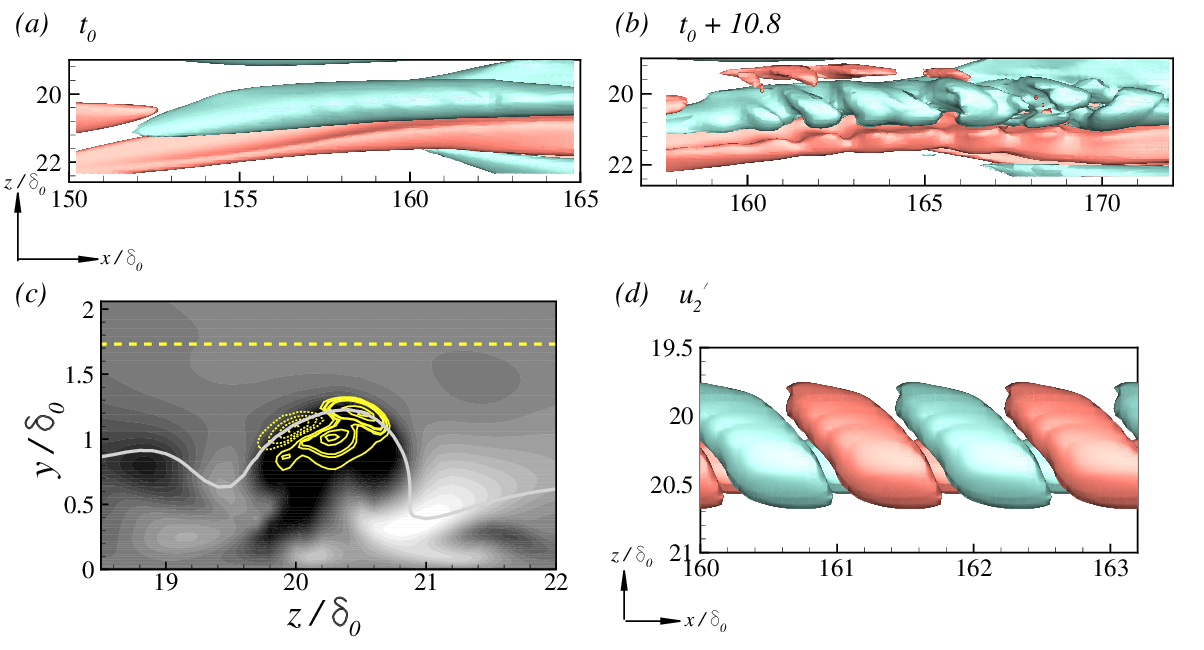}}
  \caption{Evolution of a helical instability in ZPG ($Tu=1\%$ and $A=0.5\% U_\infty$): ($a$) Isosurfaces of $u'=\pm 0.1$ at $t_0$ and ($b$) at $t_0 + 10.8$. ($c$) Contours of $-0.2 \le u' \le 0.2$ (dark for negative values) at time $t_0$ and $x/\delta_0 \sim 160$. The yellow contours show the real part of $u_2'$ for the most unstable eigenmode (dashed lines indicate negative values). The grey solid line and the horizontal yellow dashed line mark the local critical layer and $\delta_{99}$. ($d$) Top view of the iso-surfaces of $u_2' = \pm 0.025$ (red and green represent positive and negative perturbation, respectively) for this mode plotted over two wavelengths. }
\label{zpg}
\end{figure}

\citet{bose2016helical} also observed the helical mode in a ZPG boundary layer, exposed to free-stream turbulence
and a Tolmein-Schlichting wave. 
The stability results for this case is presented in figure \ref{zpg}. 
It can be hypothesized that an inflectional base flow is not necessary for the inception of helical instability. 

Figure~\ref{zpg}($a$) shows the base flow configuration at time $t_0$ in a case for which a helical instability 
was noticed at $t_0+10.8$ in the DNS---see figure \ref{zpg}($b$). 
The cross-stream plane in figure \ref{zpg}($c$) shows that the base streak configuration leading to the genesis of a helical mode 
is same as that previously described for APG (figure \ref{yz-plane}). 
The most unstable helical eigenfunction, \ref{zpg}($c$), is also similar to the APG case. 
The eigenmode is obtained for an input $\lambda_x/\delta_0=1.6$ which is similar to the average wavelength of the helical instability seen in figure \ref{zpg}($b$). 
The eigenmode from the stability analysis corresponds to $c \approx 0.624$, and $\omega_i \approx 0.39$. 
The phase speed and growth rates extracted from DNS at this streamwise station are $c \sim 0.63$ and $\omega_i \sim 0.01$. 
These results are also consistent with the analysis presented for APG. 
Figure~\ref{zpg}($d$) shows the top view of surfaces of iso-$u_2'$ values depicting the helical mode in ZPG. 

\section{Conclusions}\label{conc}

The fundamental question addressed in this paper is whether the helical patterns that have been seen in flow visualizations
of mixed mode transition
can be traced to an underlying instability.  That has been shown to be the case.  A new instability was found, that is 
distinct from the sinuous and varicose secondary instability
modes that have previously been cited in studies of bypass transition.  The helical secondary 
instability is peculiar to mixed mode transition---that is, when precursors consist of a combination of Klebanoff streaks
and instability waves.    Stability analyses were performed for  base states extracted
 from DNS of mixed mode transition in both APG and ZPG boundary layers.   Those base states are unstable to modes
 of a helical form.  Why these base states occur in the mixed mode regime remains an open question.

The stability analyses predict either two or three unstable modes, of which the helical, inner mode has the highest growth rate. 
Helical instability occurred significantly earlier than where helices were detected in DNS, indicating that 
it is the underlying cause. 

The growth rate predicted by the stability analysis for the helical mode is larger compared to that extracted from DNS. 
This too is evidence that the instability gives rise to the observed vortical structures.
The initial disturbance grows rapidly, but has small amplitude. Thus, it is swamped by higher amplitude features seen in the DNS.
As the energy of the helical mode becomes larger than the primary instability, 
helices appear in the DNS flow fields, and  quickly break down to turbulence. 


\appendix
\section{Unstable modes}\label{appA}

The stability analysis predicts three unstable modes as the mixed mode precursor shown in figure \ref{u-iso} develops downstream (figure \ref{track}). 
The eigenfunction of the most unstable helical mode (mode 1) is shown in figures \ref{yz-plane} and \ref{eigfunc}. 

\begin{figure}
  \centerline{\includegraphics[trim=0 0 0 0,clip,width=1\textwidth]{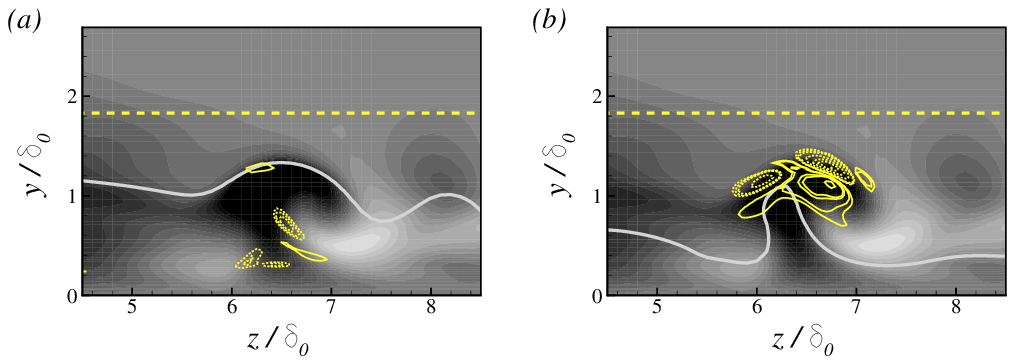}}
    \caption{Contours of real part of the streamwise perturbation eigenfunction (dashed lines indicate negative values) for the ($a$) second and ($b$) third most unstable modes at time $t_0 + 16.2$ as in figure \ref{yz-plane}($b$) at the streamwise location $x/\delta_0 \sim 185.8$. The grey solid line and the horizontal yellow dashed line mark the local critical layer and $\delta_{99}$, respectively.}
\label{mode-23}
\end{figure}

Figure~\ref{mode-23} shows the in-plane structure of the eigenfunctions of modes 2 and 3 in the same cross-stream ($y, z$)--plane and at the same time instant as in figure \ref{yz-plane}. 
The real part of $u_2'$ is shown as in figure \ref{yz-plane}($b$) for the helical mode. 
Mode 2 is activate around the stem of the mushroom-like low-speed streaky mixed mode precursor. 
Mode 3 on the other hand is also very similar to the helical mode, although, with a relatively deeper critical layer. 
The phase speeds of mode 2 and 3 are 0.558 and 0.27, respectively. 

\begin{figure}
  \centerline{\includegraphics[trim=0 0 0 0,clip,width=1\textwidth]{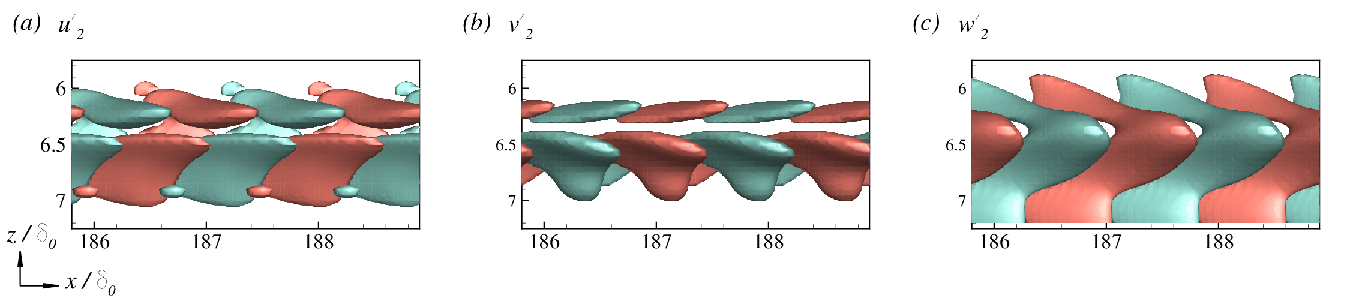}}
    \caption{Top view of the iso-surfaces of perturbation eigenfunctions (red and green represent positive and negative perturbation, respectively) for the second most unstable mode from stability analysis: ($a$) $u_2' = \pm 0.0075$; ($b$) $v_2' = \pm 0.005$ ($c$) $w_2' = \pm 0.006$.}
\label{eigfunc-2}
\end{figure}
\begin{figure}
  \centerline{\includegraphics[trim=0 0 0 0,clip,width=1\textwidth]{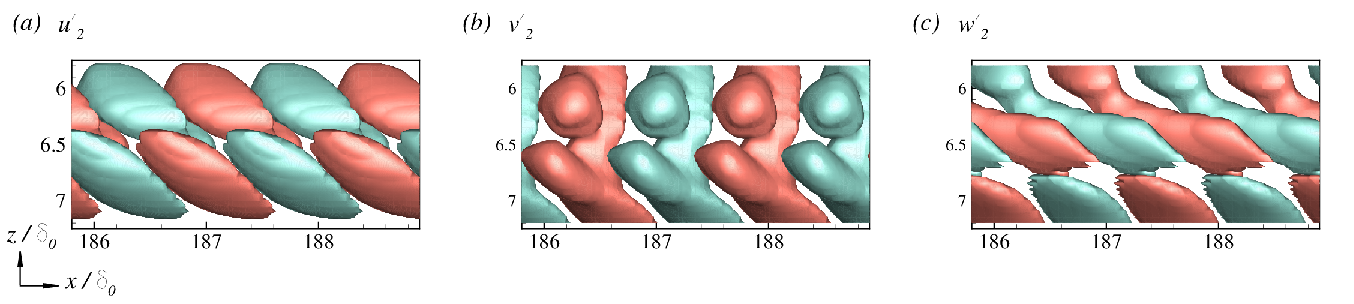}}
    \caption{Top view of the iso-surfaces of perturbation eigenfunctions (red and green represent positive and negative perturbation, respectively) for the third most unstable mode from stability analysis: ($a$) $u_2' = \pm 0.012$; ($b$) $v_2' = \pm 0.01$ ($c$) $w_2' = \pm 0.012$.}
\label{eigfunc-3}
\end{figure}
The three-dimensional structure of modes 2 and 3 are presented over two streamwise wavelength in figures \ref{eigfunc-2} and \ref{eigfunc-3}. 
Mode 2 has a symmetrical appearance. 
Mode 3 on the other hand is  helical in appearance. 
The three-dimensional structure of $u_2'$ for this mode appears to be similar to that of a helix with a higher azimuthal wavenumber than mode 1. 
$v_2'$ and $w_2'$ appears to be symmetric w.r.t. the base streaky structure. 
In the DNS, the flow structure is likely to be a superposition of the three unstable modes, although the helical mode dominates over a long time (figure \ref{u-iso}) due to its much higher growth rate.

\begin{acknowledgments}
Part of the work was performed by R.B. while supported by the NIST director's postdoctoral fellowship, which is gratefully acknowledged. 
\end{acknowledgments}


\pagebreak
\bibliography{reference}
\end{document}